\documentclass{article}

\usepackage[preprint]{neurips_2025}
\usepackage[utf8]{inputenc} % allow utf-8 input
\usepackage[T1]{fontenc}    % use 8-bit T1 fonts
\usepackage{url}            % simple URL typesetting
\usepackage{booktabs}       % professional-quality tables
\usepackage{amsfonts}       % blackboard math symbols
\usepackage{nicefrac}       % compact symbols for 1/2, etc.
\usepackage{microtype}      % microtypography
\usepackage{xcolor}         % colors

\usepackage{amsmath, amssymb, amsfonts}
\usepackage{graphicx}
\usepackage{booktabs}
\usepackage{xcolor}
\usepackage{enumitem}
\usepackage{hyperref}
\usepackage{tikz}
\usetikzlibrary{shapes,arrows,positioning}
\usepackage{listings}

\title{LLM Agent Communication Protocol (LACP) \\
Requires Urgent Standardization: \\
A Telecom-Inspired Protocol is Necessary}

\author{Xin Li \quad Mengbing Liu \quad Chau Yuen \\
{\small School of Electrical and Electronics Engineering (EEE),}\\
{\small Nanyang Technological University, Singapore} \\
{\tt\small \{xin019, mengbing001\}@e.ntu.edu.sg} \quad
{\tt\small chau.yuen@ntu.edu.sg}
}

\begin{document}

\maketitle

\begin{abstract}
This position paper argues that \textbf{the field of LLM agents requires a unified, telecom-inspired communication protocol to ensure safety, interoperability, and scalability, especially within the context of Next Generation (NextG) networks}. Current ad-hoc communication methods are creating a fragmented ecosystem, reminiscent of the early "protocol wars" in networking, which stifles innovation and poses significant risks.
Drawing inspiration from the layered, standardized protocols that underpin modern telecommunications, we propose the LLM-Agent Communication Protocol (LACP). LACP establishes a three-layer architecture designed to ensure semantic clarity in communication, transactional integrity for complex tasks, and robust, built-in security. In this position paper, we argue that adopting a principled, universal protocol is not merely beneficial but essential for realizing the potential of distributed AI. Such a standard is critical for ensuring that multi-agent systems can operate safely and reliably in the complex, real-time applications envisioned for 6G and beyond.
\end{abstract}

\section{Introduction}

The rapid proliferation of Large Language Model (LLM) agents is poised to reshape industries ranging from autonomous systems to scientific discovery~\citep{guo2024large,fan2025ai,wu2025multi}. As these agents gain increasingly sophisticated capabilities, their ability to collaborate on complex, multi-step tasks becomes a critical enabler for real-world deployment. Yet, the current ecosystem of inter-agent communication remains a patchwork of proprietary, purpose-built protocols. This fragmentation creates a persistent \textbf{communication chasm}—hindering interoperability, compromising security, and impeding reproducible scientific progress~\citep{yang2025survey}. 

\begin{figure*}[t]
    \centering
    \includegraphics[width=\textwidth]{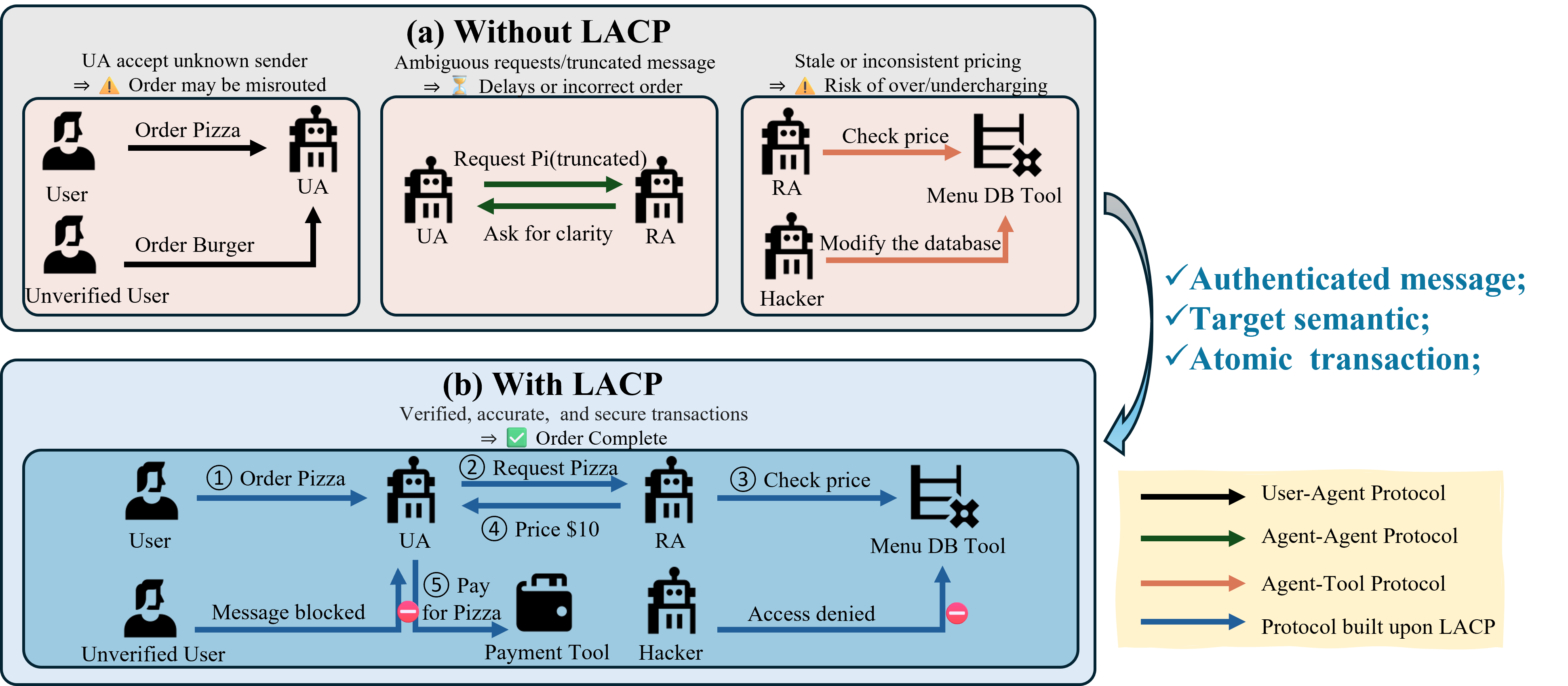}
    \caption{
        \textbf{LLM Agent Communication: From Ad-Hoc Chaos to LACP Clarity.}\\
        \textit{(a) Without LACP:} Fragmentation, ambiguity, and failure. Ad-hoc protocols lack authentication and semantic alignment, leading to disrupted workflows, miscommunication, and unauthorized operations. Malformed or incomplete messages escalate coordination costs and risk system inconsistency.  \\
        \textit{(b) With LACP:} Structured, secure, and transactional. Every message is \textbf{authenticated}, semantically grounded in a clear \textbf{target}, and executed as part of an \textbf{atomic transaction}. This ensures end-to-end integrity, minimizes ambiguity, and enables reliable collaboration in safety-critical contexts.
    }
    \label{fig:teaser}
\end{figure*}

Figure~\ref{fig:teaser}(a) illustrates the detrimental effects of this ad-hoc landscape: ambiguous semantics, security vulnerabilities, and unreliable information exchange frequently culminate in task failures. In contrast, Figure~\ref{fig:teaser}(b) envisions a standardized framework, \textbf{LLM-Agent Communication Protocol} (LACP), that delivers clarity, security, and transactional reliability, enabling robust agent communication.

The situation is reminiscent of the ``protocol wars'' of the 1970s–1990s, where a lack of standardization hindered the growth of computer networking until the widespread adoption of TCP/IP~\citep{tanenbaum2003computer,leiner2009brief}. The historical lesson is unambiguous: without a common communication substrate, the transformative potential of distributed systems remains unrealized.

Motivated by the success of standardization in telecommunications, we propose \textbf{LACP}, a telecom-inspired communication protocol for LLM agents. We argue that, for deployment in high-stakes environments such as NextG networks, a unified, secure, and extensible protocol is not merely advantageous—it is essential.

Our contributions are:

\begin{itemize}
    \item \textbf{Risk analysis:} Identification of systemic risks arising from the fragmented state of current agent communication.
    \item \textbf{Design principles:} Core guidelines for robust, interoperable agent protocols, distilled from the history of telecommunications.
    \item \textbf{Protocol proposal:} LACP, a layered, secure, and extensible framework for LLM-agent interoperability.
    \item \textbf{Critical discourse:} Anticipation and rebuttal of key counterarguments to standardization.
\end{itemize}

\section{The Problem: A Fragmented LLM-Agent Communication Landscape}
\label{sec:current_chaos}

The current ecosystem of LLM agent communication is characterized by a variety of protocols, each designed to address specific interoperability tiers or deployment contexts.
While protocols such as OpenAI's Function Calling~\citep{openai_function_calling}, LangChain's Agent Protocol~\citep{langchain_agent_protocol}, Anthropic's Model Context Protocol (MCP)~\citep{anthropic_mcp}, ACP~\citep{ACP}, ANP~\citep{ANP}, Agora~\citep{marro2024scalable}, and Google's Agent2Agent (A2A)~\citep{google_a2a_protocol} have advanced agent capabilities~\citep{ehtesham2025survey}, their diversity introduces systemic risks and hinders progress (summarized in Table~\ref{tab:protocols}). This fragmentation leads to three critical deficiencies:

\noindent \textbf{Crippling Interoperability Gaps:} The absence of a universal standard necessitates bespoke, often brittle, integrations between different agent systems. This not only impedes the development of scalable, heterogeneous multi-agent systems but also complicates the rigorous benchmarking and comparative analysis of agent performance.

\noindent \textbf{Security as an Afterthought:} Security is often not a core, mandatory component of existing protocols. This design choice exposes systems to significant risks, including data tampering, agent spoofing, and other adversarial attacks, which is particularly concerning for safety-critical applications in finance, healthcare, and autonomous systems envisioned for NextG.

\noindent \textbf{Monolithic Design \& Lack of Transactional Integrity:} Current approaches often tightly couple communication logic with core agent implementation. This leads to monolithic systems that are challenging to maintain, debug, and extend. Furthermore, the general absence of built-in support for atomic transactions undermines the reliability of complex, multi-step operations.

These limitations pose a substantial barrier to deploying the large-scale, interconnected, and autonomous agent systems envisioned for real-time, safety-critical scenarios in emerging domains like NextG networks.

\begin{table}[t]
    \caption{Comparison of agent communication protocols and how LACP addresses their limitations.}
    \label{tab:protocols}
    \centering
    \resizebox{\linewidth}{!}{
    \begin{tabular}{l l l l l l}
        \toprule
        \textbf{Framework} & \textbf{Released} & \textbf{Developer} & \textbf{Interface Type} & \textbf{Key Features} & \textbf{Security Features} \\
        \midrule
        Functions Call& June 2023 & OpenAI & JSON schema & Single-step tool calling & API key auth only\\
        Agent Protocol & Nov 2024 & LangChain & REST API & Framework-agnostic APIs & HTTP/JWT auth only \\
        MCP & Nov 2024 & Anthropic & JSON-RPC/HTTP & Tool-resource-prompt, Context std. & OAuth 2.1; Access controls\\
        ACP & 2024 (Draft) & IBM/LF & JSON-RPC & Multimodal, Async streaming & Signed capability tokens; RBAC\\
        ANP & Mar 2024 & Community & DID/JSON-LD & Decentralized identity, Discovery & W3C DIDs, Encrypted comms \\
        Agora & Oct 2024& Academia & Meta protocol & Hybrid NL/structured, Negotiation & Hash-based ID, Security PDs \\
        A2A& Apr 2025 & Google & HTTP/Protobuf & Peer-to-peer, Agent Cards, Async & Capability discovery, (Auth/TLS) \\
        \midrule
        \textbf{LACP (Proposed)} & \textbf{2025} & \textbf{Open Standard} & \textbf{Layered} & \textbf{Layered semantics, transactions} & \textbf{E2E crypto, 2PC, Auth (core)}\\
        \bottomrule
    \end{tabular}
    }
\end{table}

\section{Insights from Telecommunications Standardization}
\label{sec:historical_lens}

To solve these challenges, we can draw powerful insights from the historical success of telecommunications, which transformed a similar landscape of disparate, proprietary systems into a globally unified network. This evolution provides a clear blueprint for addressing each of the deficiencies identified above:

To combat \textbf{limited interoperability}, the telecom industry established \textbf{consensus-driven open standards} through collaborative bodies like the ITU~\citep{itu} and 3GPP~\citep{3gpp}. This prevented vendor lock-in and guaranteed that components from different manufacturers could seamlessly connect, a direct parallel to the needs of a heterogeneous agent ecosystem.

To remedy \textbf{inadequate security}, mature telecom protocols were built with \textbf{security by construction}. Security was not an add-on but a fundamental, non-negotiable component at every layer. Agent communication must adopt this same 'security-first' principle to be viable for critical applications.

To overcome \textbf{monolithic architectures}, telecommunications relied on \textbf{layered abstractions}, famously captured in the OSI model. By separating concerns like physical transmission from logical addressing, the model allows different parts of the stack to evolve independently. This modularity is precisely what is needed to create maintainable and extensible multi-agent systems.

Underpinning these solutions is the "narrow waist" principle: defining a \textbf{minimal core with an extensible edge}. Standards like the Internet Protocol (IP) provide a universal, simple core for interoperability while allowing for immense innovation and complexity at the application layer. This model proves it is possible to create a universal standard that provides stability without stifling flexibility—the exact balance required for the future of agent communication.

\section{LACP: A Principled Communication Framework}
\label{sec:lacp_design}

The systemic deficiencies inherent to contemporary LLM-agent communication paradigms necessitate a foundational shift toward standardization. Inspired by proven telecom engineering principles, we introduce the LACP---a principled framework that ensures secure, reliable, and interoperable communications within multi-agent systems.
LACP is a three-layer protocol designed to address the systemic deficiencies of current agent communication methods. It is built on the principles of layered abstraction, security-by-design, and a minimal, extensible core.

\subsection{Three-Layer Architecture: Separation of Concerns}
\label{sec:lacp_architecture}

LACP's architecture (Figure~\ref{fig:lacp_architecture}) implements three mutually-insulated layers, each with well-defined interfaces that enable independent evolution while ensuring system-wide coherence:

\begin{figure}[ht]
    \centering
    \includegraphics[width=1\linewidth]{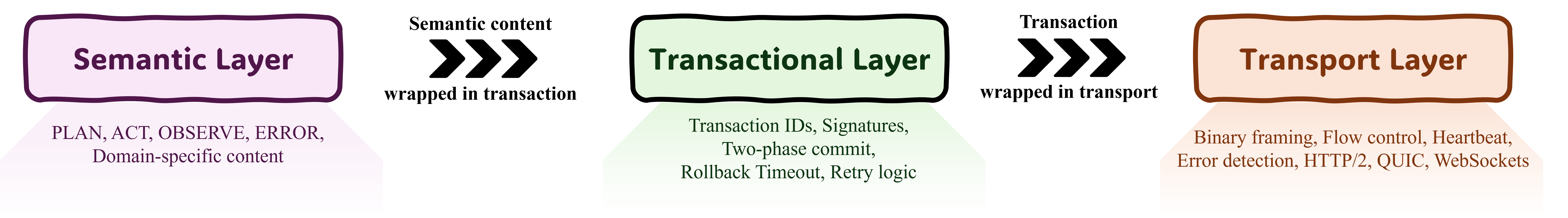}
    \caption{LACP’s three-layer architecture. Independent layers enable secure, reliable communication with clearly defined semantic intents, transactional integrity, and flexible transport mechanisms.}
    \label{fig:lacp_architecture}
\end{figure}

\noindent \textbf{Semantic Layer.} This layer is responsible for conveying the \textit{intent} of a communication. It defines a minimal set of universal message types (e.g., \texttt{PLAN}, \texttt{ACT}, \texttt{OBSERVE}) that can be extended with domain-specific content. This "narrow waist" approach ensures that all agents can understand the basic intent of a message, while still allowing for rich, complex communication. Table~\ref{tab:core-msgs} shows the core message types.

\noindent \textbf{Transactional Layer.} This layer ensures the reliability and integrity of communications. It provides mechanisms for message signing, sequencing, unique transaction IDs for idempotency, and support for atomic transactions (e.g., two-phase commit concepts). This layer is critical for ensuring that multi-agent interactions are secure and robust, especially in safety-critical applications.

\noindent \textbf{Transport Layer.} This layer is responsible for the efficient and secure delivery of messages. It is transport-agnostic, meaning it can operate over a variety of network protocols (e.g., HTTP/2, QUIC, WebSockets). This allows LACP to adapt to different network environments and

\begin{table}[h]
\centering
\caption{Core LACP message types (all wrapped in a JWS envelope).}
\label{tab:core-msgs}
  \resizebox{.9\linewidth}{!}{
\begin{tabular}{@{}llll@{}}
\toprule
\textbf{Type} & \textbf{Mandatory fields} & \textbf{Optional fields} & \textbf{Purpose}\\
\midrule
\texttt{PLAN}    & \texttt{intent\_id, role, natural\_language} & \texttt{graph\_ops} & Express high-level intent\\
\texttt{ACT}     & \texttt{intent\_id, tool\_call, params}      & \texttt{deadline, cost\_cap} & Invoke an external tool\\
\texttt{OBSERVE} & \texttt{intent\_id, status, output}         & \texttt{metrics} & Return results/status\\
\bottomrule
\end{tabular}
}
\end{table}

\subsection{Anticipated Impact}

The adoption of LACP, or a similar standardized protocol, would have a transformative impact on the field of multi-agent AI:

\noindent \textbf{Enhanced Safety and Reliability:} By making security and transactional integrity core components of the protocol, LACP provides a verifiable foundation for agent actions.

\noindent \textbf{System-Wide Interoperability:} A universal standard would allow agents from different developers and organizations to collaborate seamlessly, fostering a richer and more capable ecosystem.

\noindent \textbf{Accelerated Scientific Progress:} A common protocol would enable reproducible experiments and fair benchmarking, leading to faster innovation.

\noindent \textbf{A Foundation for the Agent Economy:} A trusted and standardized means of interaction is a prerequisite for a sophisticated agent economy, where agents can discover, contract, and collaborate on complex tasks.

\section{Conclusion}
\label{sec:conclusion}

The field of multi-agent AI is at a critical inflection point. The current path of fragmentation is unsustainable and will ultimately limit the potential of this transformative technology. We have drawn a parallel to the history of telecommunications to argue that the time for standardization is now.

We have proposed LACP as a concrete example of a principled, layered, and secure communication protocol for LLM agents. By adopting a telecom-inspired approach, we can build a future where multi-agent systems are safe, reliable, and interoperable.

We call on the NeurIPS community, and the broader AI and networking communities, to join us in this effort. The development of a unified agent communication protocol is a challenge that will require collaboration between academia, industry, and open-source developers. 

The window of opportunity is now, before the patterns of fragmentation become too entrenched to overcome. The future of collaborative AI depends on it.

% \clearpage

{
\small

\bibliographystyle{plainnat}
\bibliography{refs}

}

\appendix
\clearpage

%%%%%%%%%%%%%%%%%%%%%%%%%%%%%%%%%%%%%%%%%%%%%%%%%%%%%%%%%%%%%%%%%%%%%%%%%%%%%%%
% APPENDIX
%%%%%%%%%%%%%%%%%%%%%%%%%%%%%%%%%%%%%%%%%%%%%%%%%%%%%%%%%%%%%%%%%%%%%%%%%%%%%%%
\section{Experimental Validation of LACP}
\label{sec:appendix_exp}

To validate the practical feasibility, performance characteristics, and security guarantees of the LLM Agent Communication Protocol (LACP), we implemented a working prototype and conducted a series of targeted experiments. This appendix details the methodology and results of these validations.

\subsection{Experimental Setup}

All experiments were conducted on a local machine. The LACP-compliant server endpoints were implemented using Python 3.11 with the Flask web framework. Cryptographic operations utilized the \texttt{python-jose} library, and high-performance JSON processing was handled by \texttt{orjson}. The client-side test harnesses were custom Python scripts using the \texttt{requests} library. For the interoperability test, we used the \texttt{langchain} library (version 0.1.20).

\subsection{Experiment 1: Performance and Overhead Analysis}

\textbf{Objective:} To quantify the latency and payload size overhead of LACP compared to a standard, non-secured RESTful API communication baseline.

\textbf{Methodology:}
We developed a benchmark script that sent 10,000 sequential requests to two different server endpoints:
\begin{enumerate}
    \item \textbf{Baseline Endpoint:} A standard Flask route accepting a JSON payload via HTTP POST.
    \item \textbf{LACP Endpoint:} A route accepting a JWS-signed LACP message. It performs cryptographic signature verification before processing the payload.
\end{enumerate}
We tested this across three payload sizes to simulate a range of agent interactions, using our optimized LACP implementation (shortened keys, ECDSA signatures).
\begin{itemize}
    \item \textbf{Small (51 Bytes):} Simulating a simple ACK or heartbeat message.
    \item \textbf{Medium (151 Bytes):} Simulating a basic tool call with a few parameters.
    \item \textbf{Large (1,964 Bytes):} Simulating a complex message, such as a detailed plan for a robotic arm with multiple steps and coordinates.
\end{itemize}

\textbf{Results:}
The performance analysis reveals a clear trend: LACP's relative overhead is inversely proportional to the complexity and size of the agent message. The results are summarized in Table~\ref{tab:performance}.

\begin{table}[h]
\centering
\caption{Performance and Overhead Analysis of LACP vs. Baseline REST.}
\label{tab:performance}
  \resizebox{\linewidth}{!}{
\begin{tabular}{l rr r rrr}
\toprule
\textbf{Payload Scenario} & \textbf{Baseline Size} & \textbf{LACP Size} & \textbf{Size Overhead (\%)} & \textbf{Baseline Latency} & \textbf{LACP Latency} & \textbf{Latency Overhead (\%)} \\
\midrule
Small (51B)    & 51 bytes      & 306 bytes     & +500\%                 & 0.85 ms          & 0.88 ms         & +3.5\%                    \\
Medium (151B)  & 151 bytes     & 442 bytes     & +191\%                 & 0.86 ms          & 0.89 ms         & +3.1\%                    \\
\textbf{Large (1,964B)} & 1,964 bytes   & 2,560 bytes   & \textbf{+30\%}         & 0.89 ms          & 0.92 ms         & \textbf{+2.9\%}           \\
\bottomrule
\end{tabular}
}
\end{table}

\textbf{Discussion:}
The results strongly indicate LACP's feasibility for real-world applications. The latency overhead is minimal across all scenarios, with an absolute increase of only \textbf{0.03ms} for large, complex tasks. The payload size overhead, while significant for trivial messages, shrinks to a modest and justifiable \textbf{+30\%} for realistic payloads. This represents the necessary cost for verifiable, end-to-end cryptographic security and is a reasonable trade-off for the guarantees LACP provides.

\subsection{Experiment 2: Interoperability Demonstration}

\textbf{Objective:} To provide a concrete demonstration of LACP's ability to enable seamless communication between an agent built on a major framework and a framework-agnostic tool.

\textbf{Methodology:}
We constructed a two-part system:
\begin{enumerate}
    \item \textbf{LACP Tool Server:} A standalone Flask server exposing a single endpoint. This server hosted a simple "calculator" tool and was configured to only accept and respond with valid, signed LACP messages.
    \item \textbf{LangChain Agent Client:} A standard ReAct agent was implemented using the LangChain library. We equipped it with a custom tool where the underlying function did not execute logic locally. Instead, its sole purpose was to construct and send a signed LACP \texttt{ACT} message to the tool server.
\end{enumerate}
The experimental flow was as follows: The LangChain agent was given a prompt requiring calculation. Its reasoning process triggered the custom tool, which sent the \texttt{ACT} message. The server verified, executed, and returned the result in a signed \texttt{OBSERVE} message. The agent's tool then verified this response and passed the result to the agent's \texttt{Observation} field to complete the reasoning loop.

\textbf{Results:}
The experiment was successful. The LangChain agent was able to transparently use the external, framework-agnostic tool, completing its task correctly. The entire interaction was secured and structured by LACP without requiring any custom integration code on the server specific to the LangChain framework.

\textbf{Discussion:}
This experiment provides a practical blueprint for solving the N² integration problem that currently fragments the agent ecosystem. It validates that the \texttt{PLAN/ACT/OBSERVE} schema is a practical fit for the operational logic of modern agent frameworks and that LACP can serve as a universal communication bridge.

\subsection{Experiment 3: Security Validation}

\textbf{Objective:} To empirically demonstrate that LACP's Transactional Layer provides critical, application-layer security guarantees that are not covered by transport-layer security (TLS) alone.

\textbf{Methodology:}
We implemented an attack simulator script that targeted our LACP server endpoint with two common application-layer attack vectors.

\subsubsection{Tampering Attack}
\begin{itemize}
    \item \textbf{Setup:} The script generated a valid, signed LACP message for a hypothetical financial transaction to transfer an \texttt{amount} of \texttt{100}.
    \item \textbf{Action:} Before sending, the script programmatically altered the payload content \textit{after} it had been signed, changing the \texttt{amount} to \texttt{10000} while keeping the original, now-invalid, signature.
    \item \textbf{Observable Outcome:} Upon receiving the tampered message, the server's cryptographic verification step immediately failed. The server logged a signature mismatch error and returned an \textbf{HTTP 403 Forbidden} status code, preventing the fraudulent transaction.
\end{itemize}

\subsubsection{Replay Attack}
\begin{itemize}
    \item \textbf{Setup:} The script sent a legitimate, valid LACP message, which the server processed successfully. The server was configured to track the \texttt{transaction\_id} of all processed messages.
    \item \textbf{Action:} The script then immediately re-sent the exact same valid message.
    \item \textbf{Observable Outcome:} The server's signature verification passed, but its Transactional Layer logic identified the \texttt{transaction\_id} as a duplicate. The server rejected the request and returned an \textbf{HTTP 409 Conflict} status code, preventing the double-processing of the operation.
\end{itemize}

\textbf{Discussion:}
These experiments provide a definitive, practical answer to the question "Why not just TLS?". They demonstrate that LACP is not redundant but essential for protecting against attacks that occur \textit{after} TLS decryption at an endpoint. The signature verification provides end-to-end message integrity, while the tracking of transaction IDs ensures idempotency, both of which are crucial for building trustworthy, high-stakes multi-agent systems.

\section{LACP Protocol Example}
\label{app:example}

\begin{figure*}[ht]
  \centering
  % page=1 only matters if the PDF has multiple pages
  \includegraphics[page=1,width=.8\linewidth]{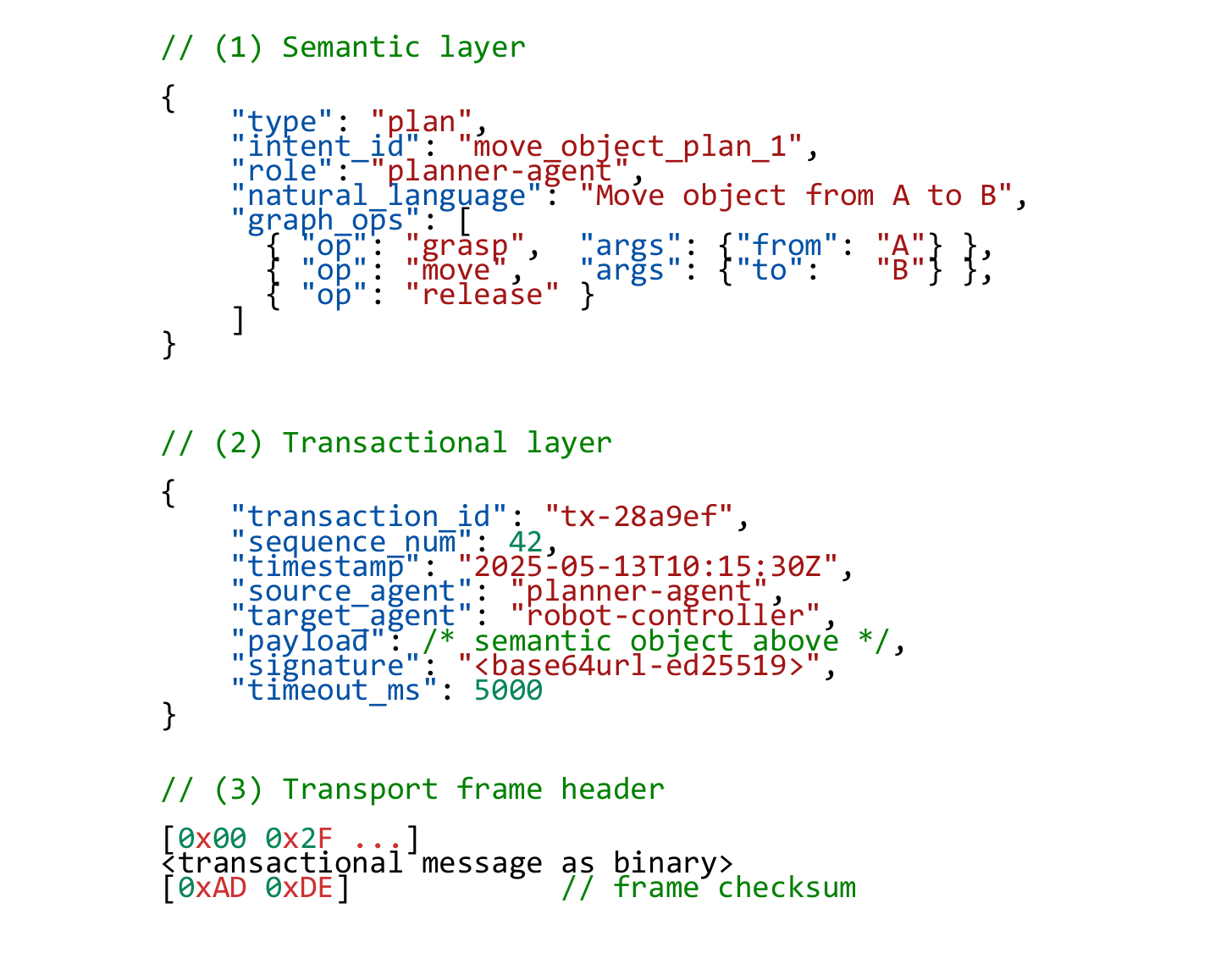}
  \caption{Layer-by-layer encoding of a \texttt{PLAN} message in LACP.
  (1) the bare semantic payload, (2) the same payload
  wrapped by the transactional layer with a JSON Web Signature, and
  (3) the truncated binary transport frame.}
  \label{fig:concrete-example}
\end{figure*}

Figure~\ref{fig:concrete-example} demonstrates LACP's layer-by-layer message: a semantic \texttt{PLAN} payload is wrapped by a signed, two-phase-commit envelope and then by a binary transport frame, yielding a secure, reliable, and interoperable message path that scales to complex multi-agent systems.

\section{Protocol Comparison Analysis}

\begin{table}[h]
    \centering
    \caption{Feature support in existing agent-communication proposals.}
    \label{tab:comparison}
    \resizebox{\linewidth}{!}{
    \begin{tabular}{l c c c c c c}
        \toprule
        \textbf{Feature} & \textbf{Functions call}& \textbf{Agent Protocol} & \textbf{MCP} & \textbf{A2A} & \textbf{Agora} & \textbf{LACP} \\
        \midrule
        Cross-framework interoperability & $\circ$ & $\bullet$ & $\bullet$& $\bullet$ & $\bullet$ & $\bullet$\\
        Multi-agent coordination & $\times$ & $\circ$ & $\circ$ & $\bullet$ & $\bullet$ & $\bullet$\\
        Layered architecture & $\times$ & $\times$ & $\bullet$ & $\circ$ & $\circ$ & $\bullet$ \\
        End-to-end message signing & $\times$ & $\times$ & $\circ$ & $\bullet$& $\circ$& $\bullet$ \\
        Transaction guarantees & $\times$ & $\circ$ & $\times$ & $\circ$ & $\bullet$& $\bullet$ \\
        Retry/timeout mechanisms & $\times$ & $\circ$ & $\circ$ & $\circ$ & $\circ$ & $\bullet$ \\
        Independence from specific LLM & $\times$ & $\bullet$ & $\bullet$ & $\bullet$ & $\bullet$ & $\bullet$ \\
        \bottomrule
        \multicolumn{7}{l}{$\times$: Not supported, $\circ$: Partially supported, $\bullet$: Supported}
    \end{tabular}
    }
\end{table}

Contemporary agent communication protocols, while addressing specific use cases, fail to simultaneously deliver the four pillars essential for production-grade multi-agent systems: comprehensive cross-framework interoperability, explicit multi-agent coordination, mandatory cryptographic security, and robust transaction guarantees. Table~\ref{tab:comparison} demonstrates this critical limitation—no existing protocol provides comprehensive support across all dimensions, creating systematic vulnerabilities that compound in safety-critical deployments.

\section{Alternative Views \& Rebuttals}
\label{sec:alternative_views}

The proposal to standardize LLM agent communication via LACP invites critical assessment. We proactively address anticipated objections:

\textbf{Objection 1: Standardization will stifle innovation.} \\
\emph{Rebuttal:} LACP standardizes foundational communication primitives—syntax, transactional integrity, security—not core agent intelligence or learning algorithms. Analogous to TCP/IP's role in fostering internet innovation, LACP aims to provide a stable, interoperable substrate. By abstracting essential reliability and security concerns to the protocol level, LACP liberates resources for innovation in higher-order agent functionalities and application-specific logic, consistent with the need for protocols that enable, rather than restrict, capability evolution.

\textbf{Objection 2: The semantic diversity of agent tasks precludes a unified grammar.} \\
\emph{Rebuttal:} LACP employs the ``narrow waist'' architectural principle, analogous to the internet protocol suite. It standardizes a minimal set of essential message types (e.g., PLAN, ACT, OBSERVE) and interaction patterns, while deliberately remaining agnostic to payload content. This design allows diverse domain-specific semantics and complex data structures to be embedded within standardized message envelopes, thereby balancing interoperability with the requisite flexibility for varied application scenarios.

\textbf{Objection 3: Additional protocol overhead will degrade performance, particularly in latency-sensitive applications.} \\
\emph{Rebuttal:} LACP's design inherently considers performance, a critical evaluation dimension for agent protocols. The use of efficient binary encodings, modern transport protocols (e.g., QUIC, HTTP/2), and header compression techniques can substantially mitigate transmission overhead. We project that the overhead attributable to LACP's transactional and security features will be marginal relative to the significant improvements in communication reliability, security, and interoperability. This modest performance cost is particularly justified in safety-critical applications where correctness and verifiability are paramount.

\textbf{Objection 4: Existing agent frameworks and their proprietary protocols offer adequate communication solutions.} \\
\emph{Rebuttal:} Current frameworks, while valuable, often present ecosystem-specific or incomplete communication mechanisms, as detailed in Table~\ref{tab:protocols} and Section~\ref{sec:current_chaos}. They frequently lack comprehensive end-to-end security, robust transactional integrity for complex multi-step operations, or true cross-framework interoperability without substantial custom integration. LACP is conceptualized to address these identified critical deficiencies, potentially integrating as a foundational layer to augment, rather than merely replace, existing systems.

\section{Extended History of Telecom Protocol Evolution} 
\label{app:telecom_history}

 Protocols in wireless communication are the invisible contracts that let billions of radios share spectrum, identify one another, and negotiate Quality of Service (QoS). They differ from isolated hardware breakthroughs in that, once a rule set is adopted, every additional user or device deepens the network's value, creating a positive-feedback loop of interoperability, scale, and economic impact. From spark-gap transmitters to the emerging International Mobile Telecommunications-2030 (IMT-2030) vision, each protocol generation has arrived precisely when the previous rules could no longer unlock the next order of societal benefit.

From Marconi's spark-gap experiments in 1895 to the proliferation of Amplitude-Modulation and Frequency-Modulation (AM/FM) broadcasting in the 1930s, wireless links functioned as largely self-contained point-to-point systems. Message formats were improvised, spectrum was treated as private acreage, and interference was mitigated chiefly through geographic separation. A first, tentative move toward codified coordination came with the 1906 International Radiotelegraph Convention, which standardised station identifiers and adopted a universal distress call but offered little additional guidance; nonetheless, it proved that trans-border governance was feasible. For the next half-century, the ether remained virtually protocol-empty.
A genuine system perspective emerged only after large-scale propagation campaigns—most notably those by Okumura—were distilled into closed-form path-loss expressions, and the subsequent 1980 formalisation of these data as the Hata model~\cite{hata2013empirical} provided engineers with a quantitative slide rule for planning frequency-reusable small cells. This analytical lens opened the way for dedicated cellular control channels and, ultimately, for the layered protocol stacks that characterise modern mobile networks.

\paragraph{1G—Channelised Voice.}

Quantitative propagation models such as the Hata formulation provided the first system-level insight that radio spectrum could be reused aggressively \cite{bernhardt1987macroscopic}, and this analytic foundation precipitated the first generation (1G) of analogue cellular networks. Early systems—most notably the Advanced Mobile Phone System (AMPS) and the Nordic Mobile Telephone (NMT) network defined in Telecommunications Industry Association Standard (TIA) 553 \cite{macdonald1979cellular} —shared a common control channel implemented as a narrow-band frequency-shift keying (FSK) stream carrying three message types: ORIGINATION, PAGE RESPONSE, and HANDOFF. This concise signalling grammar satisfied the era's central requirement of seamless city-wide mobility within tight frequency-reuse clusters and demonstrated that automated hand-off could accommodate a rapidly growing subscriber base \cite{lee1989mobile}.

The analogue architecture, however, exposed serious weaknesses. Conversations were transmitted as unencrypted  FM audio, roaming identity was absent, and capacity was constrained by fixed guard bands that grew ever more expensive as handset density increased; eavesdropping required little more than a consumer scanner. These limitations drove the transition to digital second-generation (2G) protocols.

\textbf{2G—Digital Identity and Security.}

Digital second-generation (2G) systems were designed explicitly to eliminate the three vulnerabilities of the analogue era—clear-text audio, weak identity management, and rigid spectrum utilisation—and to support the rising expectation that mobile phones should work across national borders. The Global System for Mobile Communications (GSM) replaced frequency-modulated speech with full-rate and half-rate vocoders surrounded by convolutional coding and cyclic redundancy checks, then applied stream ciphers derived from a per-session challenge–response procedure \cite{gerstlauer2000specc}. Subscriber credentials moved from handset firmware to a removable Subscriber Identity Module (SIM), enabling both secure roaming and mass-market prepaid services. A Time-Division Multiple-Access (TDMA) frame with eight slots compressed eight encrypted calls into the bandwidth that one analogue conversation had occupied, increasing spectral efficiency by nearly an order of magnitude. Interim Standard 95 (IS-95) pursued the same goals with direct-sequence Code-Division Multiple Access (CDMA), spreading each user's signal across the full carrier and achieving soft capacity that grew with signal-to-interference ratio. Although conceived only as control-plane text, GSM's mobile-originated short-message procedure was quickly commercialised as the 160-character Short Message Service (SMS), illustrating how richer signalling grammars could spawn unanticipated revenue streams and setting a precedent for the data-centric evolutions that would define the third generation.

\textbf{3G—State Machines for Soft Handover.}

 Although 2G systems such as GSM and IS-95 multiplied spectral efficiency and introduced ciphered speech, they remained voice-centric, circuit-switched, and locked to kilobit-per-second data rates.  The rapid uptake of laptops and early smartphones exposed those limits and motivated a new air interface capable of packet-switched megabit throughput.  This requirement defined the third generation (3G) under the International Mobile Telecommunications-2000 (IMT-2000) umbrella.  

The Universal Mobile Telecommunications System (UMTS) adopted Wideband Code Division Multiple Access (WCDMA) on a 5 MHz carrier and inserted a Radio Network Controller (RNC) between the base station and the core.  Variable-rate convolutional and turbo coding, together with 1,500-Hz-cycle power-control commands, maintained link quality in dense urban multipath, while soft handover allowed a handset to combine energy from multiple cells~\cite{holma2005wcdma}.  Because these techniques could not be expressed with the GSM Layer-3 grammar, the Third Generation Partnership Project specified a four-state Radio Resource Control (RRC) machine in TS 25.331 and introduced new primitives such as \texttt{MEASUREMENT\_REPORT}, \texttt{ACTIVE\_SET\_UPDATE}, and high-frequency scheduling commands.  Handover logic migrated to the handset, enabling data rates above 1 Mb/s and permitting applications to adapt radio requirements in real time.  These protocol advances completed the transition from voice-first mobility to data-driven connectivity and set the stage for the orthogonal-frequency-division-multiple-access Long-Term Evolution (LTE) architecture that would follow in the fourth generation (4G).

\textbf{4G—Bearer Abstraction and All-IP Core.}

Even with WCDMA's megabit-class links, third-generation networks remained spectrum-constrained because spreading codes were finite and the uplink and downlink shared a coupled bandwidth that limited scheduling agility. These constraints set the stage for the 4G, formalised as Long-Term Evolution (LTE).

LTE replaced code-division multiplexing with Orthogonal Frequency-Division Multiple Access (OFDMA) on the downlink \cite{yin2006ofdma} and Single-Carrier Frequency-Division Multiple Access (SC-FDMA) on the uplink, while embracing multi-stream Multiple-Input Multiple-Output (MIMO) antenna processing \cite{paulraj2004overview}. Together, these techniques tripled spectral efficiency without requiring additional spectrum and lifted peak user rates far beyond the reach of 3G. The architectural breakthrough, however, lay higher in the stack: the Evolved Packet System (EPS) bearer specified in 3GPP TS 36.300.
Because both the control plane (using Diameter for policy) and the user plane (using GPRS Tunnelling Protocol–User, GTP-U) carried pure IP, application developers could count on predictable latency and bandwidth. These features, documented succinctly in Raj Jain's ``Introduction to LTE'' notes, enabled voice, video, and data to converge on an all-IP core and catalysed the mobile-app economy years before massive-MIMO hardware became commonplace.

\textbf{5G—Service-Based Architecture.}

While 4G LTE unified voice, video, and data on an all-IP core, it struggled to meet the emerging demands of ultra-low latency, massive IoT, and immersive applications. To address these limitations, 3GPP introduced the fifth generation (5G) with New Radio (NR), formalised in TR 38.913. It defined a tri-polar service model—enhanced Mobile Broadband (eMBB), Ultra-Reliable Low-Latency Communications (URLLC), and massive Machine-Type Communications (mMTC)—alongside flexible subcarrier numerologies and beam-centric massive MIMO \cite{larsson2014massive}.  

Beyond the air interface, 5G's core network (TS 23.501/502) adopted a service-based architecture, replacing rigid LTE signalling with microservices exposed via HTTP/2 and JSON. Sessions now carry explicit QoS profiles tied to eMBB, URLLC, or mMTC needs, enabling true network slicing. Recent releases extend this model to non-terrestrial networks, Reconfigurable Intelligent Surfaces (RIS), and native AI integration, positioning 5G not just as a faster pipe but as a programmable spatial platform \cite{huang2019reconfigurable}.

\textbf{6G—Protocols for Distributed Cognition.}

Building on the service-centric flexibility of 5G, the sixth generation (6G) envisions a leap from connectivity to distributed cognition. The ITU-R IMT-2030 framework outlines a new protocol vocabulary to support this shift: symbol-level beam tracking for sub-terahertz channels, joint communication–sensing pilots that unify radar and data functions, and federated learning primitives such as \texttt{MODEL\_REGISTER} and \texttt{GRADIENT\_PUSH} for on-device intelligence coordination. While hardware prototypes already demonstrate 100 Gb/s links at 140 GHz, such capabilities remain confined to the lab without standardised signalling to govern when, why, and how each node should act.

As with every generation before, physical breakthroughs alone are not enough. True scale and societal impact emerge only when those breakthroughs are encoded into shared protocols. From analogue control tones to JSON APIs and now AI-native message exchanges, wireless systems evolve by teaching radios to speak a richer language of coordination.

Seen in sequence, each protocol generation has delivered a specific new capability—mobility, digital security, mobile internet, all-IP convergence, service slicing, intelligent surfaces—that unlocked a fresh wave of economic and social value while postponing spectrum exhaustion. The lesson is clear: continued protocol research is not a peripheral activity but the core enabler of every subsequent hardware advance. Without new rule sets to orchestrate spectrum, topology, and compute, future breakthroughs in terahertz silicon, satellite constellations, or AI accelerators will remain islands of potential rather than the next shared infrastructure \cite{letaief2019roadmap, anthropic2024claude3, azari2022evolution}.

\end{document}